\documentclass[12pt]{iopart}
\expandafter\let\csname equation*\endcsname\relax
\expandafter\let\csname endequation*\endcsname\relax
\usepackage{csquotes}
\usepackage{graphicx}
\usepackage{xcolor}
\usepackage{amsfonts}
\usepackage{comment}
\usepackage{pgfplots}
\usepackage{siunitx}
\usepackage{multirow}
\usepackage{float}
\usepackage{optidef}
\pgfplotsset{compat=newest}
\usepackage{cite}
\usepackage{hyperref}
\usepackage{harvard}

\usepackage{tikz}
\usetikzlibrary{external}
\begin{document}

\title[]{An ion treatment planning framework for inclusion of nanodosimetric ionization detail through cluster dose}

\author{Simona Facchiano$^{1,3,5}$, Ramon Ortiz$^2$,  Remo Cristoforetti$^{1,3,5}$, Naoki D-Kondo$^2$, Oliver J\"akel$^{1,3,4}$, Bruce Faddegon$^2$, Niklas Wahl$^{1,3}$}

\address{$^1$Department of Medical Physics in Radiation Oncology E040, DKFZ, Heidelberg, $^2$Department of Radiation Oncology, UCSF, San Francisco, $^3$National Center for Radiation Research in Oncology (NCRO), Heidelberg Institute for Radiation Oncology(HIRO) $^4$Department of Radiation Oncology, Heidelberg Ion Beam Therapy Center(HIT), University Hospital Heidelberg (UKHD), $^5$Faculty of Physics and Astronomy, Heidelberg University.}

\ead{simona.facchiano@dkfz-heidelberg.de}
\vspace{10pt}

\begin{abstract}
\textit{Background and aims.} 
Nanodosimetry relates the cumulative or statistical moments of Ionization Detail (ID) with biological endpoints of relevance to cancer radiotherapy using charged particles. 
This association suggests a more direct modeling of physics-based biology than established dose-response relationships and corresponding RBE-model. 
Faddegon et al. (2023) recently introduced \textit{cluster dose} $g^{(I_p)}$, as a purely physical quantity bridging the \textit{Ionization Parameter} ($I_p$) to the macroscopic treatment planning scale. In this work, we developed a framework to enable flexible and direct cluster dose optimization using a pencil-beam algorithm, which we validated with condensed history Monte Carlo (MC) simulations.

\textit{Methods.} Cluster dose combines the contributions to $I_p$ from all particles within a macroscopic volume. Our framework, implemented in the open source planning toolkit matRad \cite{wieser_development_2017}, utilizes the particle and energy-dependent $I_p$ values from an ID database precomputed from MC track strucure (MCTS) simulations. First, we create pencil-beam (PB) kernels, including fluence spectra, from condensed history MC simulations.
For a water box phantom and a representative prostate patient, we create treatment plans optimized on dose and cluster dose $F_5$ coverage and homogeneity for protons, helium and carbon ions. Plans were validated with Geant4/TOPAS MC \cite{faddegon_topas_2020,perl_topas_2012}. 

\textit{Results.} Our framework provided accurate, practical cluster dose calculation and planning. PB algorithms achieve typical accuracy for cluster dose calculation comparable to dose calculation. Recalculation with TOPAS on the box phantom yielded 3D gamma passing rates (GPRs) greater than $97\%$. For the prostate patient, GPRs exceeded $98\%$. Both used the $3\%/3mm$ criterion with a threshold of $10\%$ of the maximum dose.
Using cluster dose $F_5$ optimization, homogeneous cluster dose target coverage was achieved in all plans. A constant cluster dose prescription across all ion species shows the expected decrease in required absorbed dose for heavier ions.

\textit{Conclusions.} 
We demonstrate that fast, direct cluster dose calculation and optimization is feasible using MC validated planning with PB algorithms. Cluster dose prescription and optimization results in the expected cluster dose coverage and physical dose levels depending on the respective primary ion.
\end{abstract}

\section{Introduction}
Current radiotherapy treatment planning for protons and heavier ions relies on relative biological effectiveness (RBE) models, which translate the absorbed dose into the isoeffective photon dose. For proton therapy, a constant RBE of 1.1 is employed in clinical routines -- as recommended in the ICRU report 78 -- despite the known increase of the proton RBE towards the distal edge of a SOBP, and variations relative to different biological endpoints and patients \cite{paganetti_relative_2014,wouters_radiobiological_2015,marshall_investigating_2016}. For carbon ions, different RBE models are used for clinical purposes. One example is the local effect model (LEM), which evolved through multiple revisions \cite{scholz_calculation_1996,scholz_computation_1997,elsasser_quantification_2010}. Another is the modified microdosimetric kinetic model (MKM) \cite{hawkins_statistical_1994,kase_microdosimetric_2006,inaniwa_reformulation_2015}, from which the stochastic microdosimetric kinetic (SMK) model  has been evolving \cite{inaniwa_adaptation_2018}. The comparison between the two models shows significant deviations in the absorbed dose prescriptions of up to $15\%$.
In order to improve the biological effect in the target and healty tissue sparing, LET has been proposed as a quantity to steer in RTP. But while it is possible to reduce LET in healthy tissues without altering the dose distribution \cite{grassberger_variations_2011,giantsoudi_linear_2013,unkelbach_reoptimization_2016}, the non-linear behavior observed in multiple field treatment planning indicates that LET is not a direct predictor of biological effect \cite{grun_is_2019}. This led to the proposal of nanodosimetry as a novel approach and a logical extension of microdosimetry. 

Nanodosimetry has been gaining interest since the 1970s \cite{pszona_track_1976} with the subsequent development of experimental nanodosimetry and Monte Carlo track structure (MCTS) simulations.  Through nanodosimetry, the link has been demonstrated   between the distribution of energy depositions within nanometric volumes surrounding a particle track -- resulting in ionizations and excitations -- with complex DNA damage, such as double-strand breaks, and repairability \cite{goodhead_microdosimetry_1997,pinto_double_2002,schuemann_new_2018}.

More recently, it has been suggested to use \textit{Ionization Detail} (ID)  to condense nanodosimetric track-structure information in useful quantities. ID is defined as the spatial distribution of ionization events within nanometer sized regions surrounding a particle track \cite{ramos-mendez_fast_2018}, and characterized by the \textit{frequency ionization cluster size distribution} (fICSD). The fICSD information can be condensed into statistical moments, or cumulative quantities known as \textit{Ionization parameters} ($I_p$) \cite{ramos-mendez_fast_2018,faddegon_ionization_2023}.

Studies have confirmed the correlation between cumulative probabilities and inactivation cross sections in mammalian cells \cite{blakely_cell_1992,conte_track_2017,conte_nanodosimetry_2018,rabus_broadscale_2020,conte_track_2023}.

Therefore, the use of ID enables translation from the nanoscopic to the macroscopic scale. When successful, this enables direct treatment planning on ID statistical moments. For example, \citeasnoun{burigo_simultaneous_2019} proposed to minimize the variance of cluster frequency in addition to common RBE-based planning.

Recently, \citeasnoun{faddegon_ionization_2023} proposed a generalized paradigm for ID-based radiotherapy treatment planning (ID-based RTP). They presented a mathematical model of ID bridging the $I_p$ to the macroscopic RTP scales through the definition of a macroscopic (voxel-averaged) $I_p$ and the novel concept of \textit{cluster dose}, indicated with $g^{(I_p)}$. \citeasnoun{faddegon_ionization_2023} showed that cluster dose for specific $I_p$ can be closely associated with biological effects and become particle and energy independent, with the potential to fully accounting for LET. For example,  for $I_p$ defined as $F_k$, the number of clusters of k or more ionizations in a nanometer-sized volume of specific dimensions, $F_5$ was preferred for aerobic cells  (showing the closest association between cell survival and $F_k$ for different types of charged particles of the same fluence), while $F_7$ was preferred for hypoxic cells. Preliminary data suggested that for the preferred $I_p$, defined for its association with cell survival, the same cluster dose level is expected to lead to similar levels of cell survival. Thus the physical quantity cluster dose could be directly used in RTP without the need of biological models like RBE.

MCTS databases of nanodosimetric quantities can be employed in condensed history Monte Carlo (MC) in order to directly calculate averaged or weighted quantities on the treatment planning scales \cite{ramos-mendez_fast_2018}. Nevertheless, calculation times of MC simulations are too long to allow clinical application, although more effort has been invested into fast GPU-based MC methods \cite{tian_gpu_2015,giantsoudi_validation_2015,simoni_fred_2020}.

In this context, the purpose of this paper is twofold: (1) to show that cluster dose calculation using pencil-beam algorithm is consistent with MC calculation. (2) To introduce a generalized plan optimization approach using cluster dose.
We have implemented our framework in the open source planning toolkit matRad \cite{wieser_development_2017} and validated our tool with Geant4 condensed history MC \cite{carrier_validation_2004} in TOPAS \cite{perl_topas_2012,faddegon_topas_2020}.

Our approach is based on the ID database from \citeasnoun{faddegon_ionization_2023}. The database was  precalculated from MCTS simulations, containing particle and energy-dependent fICSD. The database can be used with particle spectra stored in the kernel database, in order to calculate the pencil-beam cluster dose. This enables cluster dose calculation -- for any definition of $I_p$ -- using the pencil-beam algorithm in matRad with subsequent and direct cluster dose plan optimization for the three available ion treatment options in matRad: protons, helium and carbon ions. 
We have created prototype plans and have evaluated their (nano-) dosimetric accuracy on a cubic water phantom and a prostate patient by comparing to a forward (cluster) dose calculation, and we have explored different absorbed dose and cluster dose prescriptions. The comparison against MC recalculation with TOPAS is provided, and results of $\gamma$-analysis \cite{low_gamma_2010} indicate the level of reliability of our ID-based RTP framework. The framework concept allows the use of custom radiation sources and nanodosimetric data.

\section{Materials and Methods}

\subsection{From Ionization detail to cluster dose: an overview of the mathematical model}\label{section1:formalism}
\citeasnoun{faddegon_ionization_2023} introduced a comprehensive formalism bridging the nanoscopic $I_p$ to the macroscopic world, using \enquote{cluster dose} as a fluence dependent target quantity for treatment planning.

They also provide a proof that the physical quantities defined within the model serve as particle-agnostic biological endpoints for cell survival.

For this reason, the aim of this research field is to include cluster dose and the $I_p$ in future radiotherapy treatment planning, in order to achieve improvement over biological uncertainties affecting currently employed models \cite{fossati_radiobiological_2018,nystrom_treatment_2020}.

The formalism relies on the previously introduced concepts of \textit{ionization cluster size} $\nu$,  defined as the number of ionizations within a nanometer-sized region, and of ICSD or fICSD $f(\nu)$, which together describe the spatial distribution of ionizations along a particle track \cite{grosswendt_nanodosimetry_2006,palmans_future_2015,burigo_distributions_2016,ramos-mendez_fast_2018}.

The dimensions of the nanometer-sized regions for scoring clusters are defined to be comparable to a DNA segment of 10 base pairs, such that two strand breaks occurring within the same volume can cause complex DNA damage, e.g. a Double Strand Break (DSB) \cite{goodhead_track_1989,michalik_model_1992,goodhead_initial_1994,grosswendt_new_2007}.

\citeasnoun{faddegon_ionization_2023} indicate fICSD as $f^c(\nu)$, meaning that it depends on the particle \enquote{class} $c$, i.e. type and energy. The $I^c_p$ reflects the condensed information of the fICSD in a single scalar or vector value, and it is particle class specific as well. While \citeasnoun{faddegon_ionization_2023} formalized the computation of $I^c_p$ as a linear operator acting on $f$, two types of $I_p$ were specifically highlighted and will be used in this manuscript:
\begin{equation}
   \hspace{0cm} F_k = \sum_{\nu = k}^{\infty} f^c(\nu) \hspace{50pt} N_k = \sum_{\nu = k}^{\infty} \nu f^c(\nu)\ .
\end{equation}
$F_k$ is the number of clusters with $k$ or more ionizations, $N_k$ is the number of ionizations contained within the $F_k$ clusters, both per unit path length.

\textit{Cluster dose} $g_j^{(I_p)}$ is the macroscopic generalization of the $I_p$ as it is the fluence-weighted sum of $I^c_p$ from all the particle classes within a macroscopic voxel
\begin{equation}
\label{clusterDoseEq}
    \hspace{0cm} g_j^{(I_p)} = \displaystyle \frac{1}{\rho_0} \sum_{c\in C_j} \phi_j^c I_p^c = \frac{1}{\rho_0} \sum_{c\in C_j} \frac{t_j^c}{V_j} I_p^c\ ,
\end{equation}
where $C_j$ is the set of particle classes within voxel $j$, $\phi_j^c$ indicates the fluence of particle class $c$ within voxel $j$, $I_p$ can be $N_k$ or $F_k$, $t_j^c$ is the track length of particle class $c$ within voxel $j$, $V_j$ the voxel volume, and $\rho_0$ is the density of the medium used to compute the database (i.e., water). Analogous to absorbed dose, that represents the deposited energy per unit mass, cluster dose represents the number of ionization clusters per unit mass, and it is therefore expressed in units of inverse mass.

It is possible to define a voxel-averaged $I_p$ computed as the fluence-averaged $I_p$ within a voxel:
\begin{equation}
\label{averagedIp}
    \hspace{0cm} I_p^{C_j} = \displaystyle \frac{\sum_{c\in C_j} \phi_j^c I_p^c}{\sum_{c\in C_j} \phi_j^c }
\end{equation}
from which it follows the relation between cluster dose, $I_p$ and fluence:
\begin{equation}
\label{eq:Ipvsg}
    \hspace{0cm} g_j^{(I_p)} = \displaystyle   \frac{1}{\rho} \phi_j  I_p^{C_j}, \hspace{1cm} with \hspace{0.5cm} \phi_j = \sum_{c\in C_j} \phi_j^c\ .
\end{equation}
Equation (\ref{eq:Ipvsg}) shows that if $g^{(I_p)}$ is analogous to dose, the averaged $I_p^{C_j}$ corresponds to stopping power and can be interpreted as a radiation quality metric. Consequently, introducing cluster dose optimization alongside the steering of the $I_p$ becomes essential for managing radiation quality effectively.

\subsection{Practical cluster dose calculation}\label{section2:clusterDoseCalc}

The MC method is currently regarded as the most accurate approach for simulating particle transport in a medium, as they sample interactions according to radiation transport equations and respective cross section, fully accounting for details of heterogeneity.  Most often, radiotherapy uses condensed history MC, which groups events to enhance computational efficiency while retaining accuracy relevant to treatment planning   \cite{paganetti_monte_2014,schuemann_assessing_2015,liang_comprehensive_2019,teoh_is_2020}. Thus, MC methods would be the primary choice for nanodosimetric calculations, yet they are rarely fully integrated into the ion planning process, especially for ions treatments. Instead, they are often used for secondary dose verification or for research applications \cite{bueno_algorithm_2013,grassberger_quantification_2014,saini_dosimetric_2017,taylor_pencil_2017,piffer_validation_2021}. 

While recently more effort has been put into creating faster MC algorithms \cite{tian_gpu_2015,giantsoudi_validation_2015,simoni_fred_2020,lysakovski_development_2024}, ion therapy treatment planning usually utilizes pencil-beam (PB) algorithms (often derivatives from \citeasnoun{hong_pencil_1996}) interpolating from a pre-computed kernel database created from MC simulations or measurements in water. Simplifying the inclusion of lateral heterogeneities and relying on water-equivalent path length, they provide the speed necessary for dose influence matrix calculations at reduced accuracy \cite{taylor_pencil_2017,liang_comprehensive_2019}. The dose influence matrix is then used to map individual beamlet fluences to the full dose distribution for changing input fluence, which is primarily necessary during iterative plan optimization.

Building on this principle, we investigated two approaches for nanodosimetric cluster dose calculation -- a scoring method embedded into MC simulation for forward (cluster) dose calculations as well as a pencil-beam kernel algorithm for inverse planning via the calculation of cluster dose influence matrices.

Section \ref{section1:formalism} shows that the required inputs for cluster dose calculation are the fICSD (or the $I_p$), and the primary and secondary particle fluence. The former can be provided as a precomputed ID database, with the general format outlined below (see \textit{Ionization Detail Database}). Particle fluence can be obtained through MC, from which cluster dose can be directly calculated within a simulation. Alternatively, fluence can be stored as fluence kernels to calculate cluster dose in post-processing and use with pencil-beam algorithm. 

\subsubsection*{Ionization Detail Database.}
The cluster dose calculation methods described in the following sections both rely on the availability of a precomputed database of ID, which contains the particle and energy-dependent fICSD (or $I_p$). The general format we describe here is inspired by the MCTS database presented by \citeasnoun{faddegon_ionization_2023}.
For each particle type $p$, an fICSD look up table is required -- or equivalently, two look up tables: one for $F_K$ and one for $N_K$. The fICSD is tabulated as a function of $\nu$ and E, while $F_k$ and $N_k$ are tabulated as a function of $k$ and E.
Given this format, the fICSD of a proton with kinetic energy E, i.e. $f^{c}(\nu)$ with $c = \{proton, E\}$ (see Section \ref{section1:formalism}), is interpolated from the proton fICSD table:
\begin{equation}\label{eq:fICSDinterp}
    \hspace{0cm} f^{c}(\nu) \approx \frac{f^p(E_j, \nu) - f^p(E_i, \nu)}{E_j-E_i} (E-E_i) + f^p(E_i,\nu), \quad \quad E_i\leq E \leq E_j
\end{equation}
and analogously for the $I_p$, i.e. $F_k$ and $N_k$.

\subsubsection{Computing cluster dose $g$ with Monte Carlo.}\label{sec:MonteCarloCDscoring} Given a precomputed \textit{ID database} with the format described, cluster dose can be directly scored within a MC simulation. According to Equation (\ref{clusterDoseEq}), cluster dose is proportional to the track-weighted sum of the $I_p$ of the particle classes within a voxel. Thus it is sufficient to cumulate the product between the particle track length and the $I_p$, the latter being interpolated on the fly using the current particle type and energy assigned to the step. This method is considered to provide the gold-standard reference value, as MC codes have previously been validated for dose calculation \cite{carrier_validation_2004,tourovsky_monte_2005,saini_dosimetric_2017,kozlowska_fluka_2019}. Additionally, the $I_p$ (or the frequencies fICSD) are interpolated for each single scored particle step. 

\subsubsection{Pencil-beam algorithm for cluster dose $g$ calculation.}\label{sec:gcalcConcept} 

The kernel database used for dose calculation using pencil-beam algorithm generally contains the depth-dose profile, representing the integrated dose at each depth, as well as the lateral beam broadening due to Multiple Coulomb Scattering. The latter is often modeled using Gaussian distributions (single, double, or multiple), with the width(s) of the distribution varying as a function of depth. Thus we extend the kernel database in order to be able to calculate the pencil-beam cluster dose.
 
We propose two methods for PB cluster dose calculation -- which are further explained below and indicated as \textit{Flexible} and \textit{Fast} methods: (1) using particle spectra and directly interpolating the $I_p$ from the ID database, and (2) using a precomputed cluster-dose kernel for the respective $I_p$. 
Thus we extend our database by including both precomputed fluence and cluster dose kernels obtained through MC simulations (see section \ref{sec:MonteCarloCDscoring}).

Regarding fluence, within a single pencil-beam and for each relevant primary and secondary particle type -- identified by the atomic number Z and, if applicable, the atomic mass A -- we store: the fluence as function of depth $\phi^{[Z,A]}(d)$, fluence as function of the energy bin and depth $\phi^{[Z,A]}(E_{bin},d)$, and the lateral beam broadening as function of depth. For the latter we consider a triple radial Gaussian model:
\begin{equation}
\hspace{0cm}\mathcal{G}_{\sigma}(\rho) \equiv \mathcal{G}(0,\sigma, \rho) = \frac{e^{-\rho^2/2\sigma^2}}{2\pi \sigma^2},
\end{equation}
\begin{equation}
\hspace{0cm}\int_0^\infty \mathcal{G}_{\sigma}(\rho) 2\pi \rho\hspace{3pt} d\rho = 1,
\end{equation}
\begin{equation}
\label{tripleGauss}
\hspace{0cm}\mathcal{L(\rho)}  \equiv (1-\omega_2-\omega_3) \cdot \mathcal{G}_{\sigma_1}(\rho) + \omega_2 \cdot \mathcal{G}_{\sigma_2}(\rho) + \omega_3 \cdot \mathcal{G}_{\sigma_3}(\rho),
\end{equation}
\begin{equation}
\hspace{0cm}\int_0^\infty \mathcal{L(\rho)} 2\pi \rho\hspace{3pt} d\rho = 1\ .
\end{equation}

Here, $\mathcal{G}_{\sigma}(\rho)$ is the single radial Gaussian distribution, and $\mathcal{L(\rho)}$ indicates the lateral fluence distribution. For each $(Z, A)$ the set $\{\sigma_1(d), \sigma_2(d), \sigma_3(d), \omega_2(d), \omega_3(d)\}$ is stored, obtained by fitting particle fluence from MC simulations of pencil-beams in water.

Regarding cluster dose, we include depth profiles $g^{(I_p)}(d)$ of pencil-beams in the kernel database. This can be achieved either by directly scoring the track-weighted sum of the $I_p$ in each voxel (voxel-averaged $I_p$) during a MC simulation, as detailed in Section \ref{sec:MonteCarloCDscoring}, or through post-processing. The latter approach requires the fluence kernel database and Equation (\ref{clusterDoseEq}) to compute the integrated cluster dose at each depth bin. 
Specifically, from Equation (\ref{clusterDoseEq}), the contribution of a particle type $s = (Z, A)$ to the cluster dose in a depth bin, including the full set of particle classes (energies) of that particle type, is expressed as the following matrix product:
\begin{equation} \label{eq:fluencexIP}
    \hspace{0cm} g^{I_p[s]}(d_j) = \frac{1}{\rho_0} \sum_{E_i} I_p^{[s]}(E_i) \hspace{2pt} \phi^{[s]}(E_i,d_j)   
\end{equation}
where $\phi^{[s]}(E_i,d_j)$ represents the fluence of particle type $s$, binned by energy and depth, and stored in the database. The depth binning is linear, with finer binning of \SI{0.1}{\milli\meter} around the peak, \SI{5}{\milli\meter} binning in the plateau region, and \SI{10}{\milli\meter} binning in the tail. The energy binning is also linear, with the bin size set to the minimum available energy in the ID database for a given particle type. The $I_p^{[s]}(E_i)$ is interpolated from the ID database for energy $E_i$. The total cluster dose in depth bin $d_j$ is then given by the sum of $g^{I_p[s]}(d_j)$ over all particle types $s$.

\paragraph{Flexible cluster dose calculation with spectral fluence kernels.}
The first approach relies on precomputed spectral fluence kernels for primary and secondary particles, allowing for on the fly interpolation of the ID database. 

We refer again to Equation (\ref{eq:fluencexIP}), which expresses the contribution of a particle type $s$ to cluster dose at a given depth. In order to express the cluster dose contribution within a bin volume at a radial distance $r$ from the beam axis, we must multiply $g^{(I_p)}_{[s]}(d_j)$ by the correspondent lateral scattering. If $\mathcal{L}^{[s]}_{d_j}(r_k)$ represents the lateral broadening of fluence for particle type $s$ at depth bin $dj$ and radial distance bin $r_k$, we approximate the total cluster dose within the same bin volume as
\begin{equation}
\label{clusterDoseCalcEq}
    \hspace{0cm} g^{(I_p)}(d_j, r_k) = \sum_{s} g^{(I_p)}_{[s]}(d_j, r_k) \approx  \sum_{s}\mathcal{L}^{[s]}_{d_j}(r_k) \hspace{5pt} g^{(I_p)}_{[s]}(d_j)\ .
\end{equation}

Here we approximate the lateral broadening of cluster dose using that of the fluence, assuming depth dependence and neglecting the effects of a change in lateral energy distributions. This choice is motivated by our aim to use the lateral broadening separated by particle type, which is directly available for fluence from MC simulations without any post-processing modifications. 

\paragraph{Fast cluster dose calculation from precomputed cluster dose kernels.}
The second approach targets less computational effort by precomputing cluster dose kernels, forgoing the flexibility of exchanging or modifying the ID database.
Total cluster dose depth profiles $g^{(I_p)}(d_j)$ are stored in the kernel database (similarly to dose kernels), accounting for all primary and secondary particles available within the database and contributing to the cluster dose. In this approach, the lateral scattering modeling for the absorbed dose kernels is assumed to approximate the cluster dose lateral broadening.

The total cluster dose at depth bin $d_j$ and radial distance $r_k$ is then expressed as:
\begin{equation}
    \hspace{0cm} g^{(I_p)}(d_j, r_k) \approx g^{(I_p)}(d_j)\hspace{5pt}\mathcal{L}^{dose}_{d_j}(r_k)\ . 
\end{equation}
Here, $\mathcal{L}^{dose}_{d_j}(r_k)$ indicates the absorbed dose lateral broadening, modelled as a single or double Gaussian.

\subsection{Cluster dose optimization} \label{section3:clusterDoseOptiProblem}
One of the main goals of this work is to be able to directly optimize with prescriptions on cluster dose.
Thus simultaneous optimization of (RBE) dose and cluster dose can be described by the following problem:

\begin{mini}
{\mathbf{w}}{\mathbf{\chi}(\mathbf{d, g}) = \sum\limits_{n = 1}^N  \left(p_{n, d} f_{n}(\mathbf{d}) + p_{n, g} h_{n}(\mathbf{g}) \right)}
{\label{eq:optiProbNew}}{}
\addConstraint{\mathbf{d} \in U}{ = \cap_n U_n }{U_n =  \{\mathbf{d} | d^{\min} < d_i < d^{\max}, \quad \forall i \in J_n \}}
\addConstraint{\mathbf{g} \in V}{ = \cap_n V_n }{V_n =  \{\mathbf{g} | g^{\min} < g_i < g^{\max}, \quad \forall i \in J_n \}}
\addConstraint{d_i=}{\sum_j D_{ij} w_j}{\quad \quad \forall i}
\addConstraint{g_i=}{\sum_j G_{ij} w_j}{\quad \quad \forall i}
\addConstraint{w_i\geq}{ 0}{\quad \quad \forall i}
\end{mini}

where N is the number of structures within the RTP geometry, e.g., target volumes and OARs. The beamlets fluence weights are indicated by vector $\mathbf{w}$. The associated dose and cluster dose objective functions are indicated as $f_{n}(\mathbf{d})$ and $h_{n}(\mathbf{g})$ respectively, with $p_{n, d}$ and $p_{n, g}$ the corresponding penalties. The set of constraints acting on absorbed dose and cluster dose are indicated by $U$ and $V$, with $J_n$ indicating the voxel indices for structure $n$.

As cluster dose can now be expressed as a linear transformation of fluence to cluster dose via the cluster dose influence matrix ($G_{i,j}$), we chain dosimetric prescriptions on cluster dose distribution to the fluence analogously to dose, with respective considerations for the derivatives. This enables us to use the same objective and constraint functions as for conventional dose-based treatment planning, but instead acting on cluster dose. In this work we use three objective functions, i.e. least-squares and variance minimization to achieve homogenous cluster dose in targets, and squared overdosing for OARs. A more explicit form for the objective function is: 

\begin{equation} \label{eq:objectivefuncGeneral}
    \hspace{-1.5cm}
    \begin{array}{rcl}
        \chi(\mathbf{d, g}) & = & \displaystyle\sum_{n = 1}^{N_{TARGET}} p_{n, d}^T \left(d_i - d_n^{ref} \right)^2 
        + \displaystyle\sum_{n = 1}^{N_{OAR}} p_{n, d}^O \left(d_i - d_n^{ref} \right)^2 \theta(d_i - d_n^{ref}) \\
        
        & + & \displaystyle\sum_{n = 1}^{N_{TARGET}} p_{n, g}^T \left(g_i - \overline{g} \right)^2 
        + \displaystyle\sum_{n = 1}^{N_{OAR}} p_{n, g}^O \left(g_i - g_n^{ref} \right)^2 \theta(g_i - g_n^{ref})
    \end{array}
\end{equation}

where the first line of the Equation represents the objective function relative to dose, and the second line is relative to cluster dose. Here, $d_i = d_i(\mathbf{w})$ and $g_i = g_i(\mathbf{w})$ indicate respectively dose and cluster dose for single voxel $i$. The target objective contains $\overline{g}$, that is the average cluster dose $g(\mathbf{w})$ over all the target volume.

The choice of the penalties associated to the objectives can determine the priority quantity within the optimization, and it is similar to the common formulation of the conventional treatment planning problem.

\subsection{Implementation in matRad: a workflow for cluster dose planning} \label{section4:matRadImplementation}
Dose calculation and cluster-dose optimization as described in sections \ref{section3:clusterDoseOptiProblem} was implemented in matRad, an open-source software for radiotherapy treatment planning with photons and ions \cite{cisternas_matrad_2015,wieser_development_2017}. While matRad's pencil-beam algorithm was extended to compute cluster-dose, the TOPAS MC interface was used together with custom cluster dose scorers, and the optimization routine was extended to allow cluster dose as an additional optimization routine.

The ID-database we used is an MCTS database computed according to the formalism presented in \citeasnoun{faddegon_ionization_2023} and provided by collaborators.

For dose calculation, we precalculated general-purpose (which means not resembling an existing machine) kernel databases for protons, helium and carbon through MC in TOPAS using OpenTOPAS v.4.0. They contain absorbed dose kernels with the lateral broadening modeled by a double radial Gaussian distribution. 
 
Particle fluence is included, as well as the triple radial Gaussian distributions modeling the lateral fluence broadening for each particle type. The energy of particles used to build the fluence spectra is the one at the beginning of the step\footnote{The option to bin fluence by the initial energy in a MC step is indicated with "Prestep" in TOPAS.} in TOPAS simulations.
The sets of particle types considered in our database are (See also Table \ref{tab:particleTypes}):
\begin{description}
\item[a) protons:] the database for protons includes hydrogen $(Z, A) = (1, 1)$, deuterium $(Z, A) =(1, 2)$, tritium $(Z, A) =(1, 3)$, helium $Z=2$, and oxygen $Z=8$.

\item[b) helium ions:] the helium database includes the same particle types as for protons.

\item[c) carbon ions:] the carbon database includes all the particle types with Z up to 8, which means the same as the proton and helium basedata, plus lithium $Z=3$, beryllium $Z=4$, boron $Z=5$, carbon $Z=6$, nitrogen $Z=7$, and oxygen $Z=8$.
\end{description}
\begin{table}
\caption{\label{tab:particleTypes} Summary of particle types included in fluence kernel basedata according to primary particle.}
\footnotesize
\begin{tabular}{p{2.8cm} | p{1.5cm} p{1.5cm} | p{1.5cm} p{1.5cm} | p{1.5cm} p{1.5cm}}
\br
Primary particle & \multicolumn{2}{c|}{Protons} & \multicolumn{2}{c|}{Helium} & \multicolumn{2}{c}{Carbon} \\
\mr
\multirow{3}{*}{\parbox{2.5cm}{Particle types \\ included in \\ fluence spectra}}& Z & A & Z & A & Z & A \\
\cline{2-7}
& 1 & $\left\{1,2,3\right\}$ & 1 & $\left\{1,2,3\right\}$ & 1 & $\left\{1,2,3\right\}$ \\
& $\left\{2,8\right\}$ & any  & $\left\{2,8\right\}$ & any & $\left\{2, ...,8\right\}$ & any \\
\br
\end{tabular}
\end{table}

Through post-processing of the fluence kernels, we included, for each pencil-beam, the total cluster dose depth profiles $g^{(I_p)}(d)$ precomputed in water for all relevant $I_p$ , $\forall I_p \in  \{F_k, N_k| k \in [1, ..., 10]$\}.

Both cluster dose calculation methods described in section \ref{sec:gcalcConcept} were implemented. In the flexible method, matRad reads the particle fluence from the kernel database and interpolates the correspondent $I_p$ on the fly, from the MCTS database. In the fast method, precomputed $g^{(I_p)}$ depth profiles are used along with the absorbed dose lateral broadening.

The optimization problem (See Section \ref{section3:clusterDoseOptiProblem}) was also implemented: a new set of objectives and constraints, acting directly on cluster dose, was added to matRad. This enables direct steering of cluster dose in radiotherapy treatment planning in matRad.  

\subsection{Treatment planning study on water box and prostate patient} \label{section5:phantomsApplications}

To demonstrate cluster dose treatment planning and validate the kernel algorithm against MC simulations, we compare a set of treatment plans planned with the pencil-beam algorithm in matRad to recalculated MC (cluster) dose distributions with matRad's TOPAS interface. To demonstrate the impact and successful steering of cluster dose in optimization, we explore different prescriptions, phantom configurations, and radiation modalities.

For our study, we chose to use $F_5$, the preferred $I_p$, among $F_k$ $\forall k=1, ...,10$,  for normoxic cells, observed for cell survival for two cell lines, for ions from protons to argon, by \citeasnoun{faddegon_ionization_2023}, to calculate cluster dose $g^{(F_5)}$. In order to improve calculation time, we used the fast method presented in Section (\ref{sec:gcalcConcept}) employing precomputed $g$-depth profiles and approximating the lateral broadening of cluster dose with the lateral broadening of absorbed dose. We also added results from calculation of cluster dose using the flexible method reyling on fluence-spectra kernels (Section \ref{sec:flexiCalc}) and different $I_p$ (Section \ref{sec:multipleIp}).

\subsubsection{Box phantom}
First, we evaluated cluster dose calculation and optimization on a simple cubic water phantom. The target is a cube of \SI{60}{\milli\meter} width, located in the center of a larger water cube of \SI{480}{\milli\meter} width that is divided into $160^3$ cubic voxels of \SI{3}{\milli\meter} edge each. We simulated irradiation with one field, separately exploring a constant absorbed dose SOBP prescription, $D = \SI{2}{Gy}$, and a constant cluster dose SOBP of $g^{(F_5)} = \SI{1.3}{\per\pico\gram}$ in the target volume, for protons, helium and carbon ions. There were no constraints applied. For each field, $10^8$ primary particles were simulated for MC validation.

\subsubsection{Prostate patient}

Subsequently, we create several plans with different prescriptions on a prostate patient. We simulated irradiation target PTV prostate with two opposing lateral fields. For each of the three ions, we again explored an absorbed dose prescription $D = \SI{2.27}{\gray}$, and cluster dose prescription $g^{(F_5)} = \SI{1.3}{\per\pico\gram}$ in the prostate PTV. There were no constraints applied. MC simulations used $10^8$ primary particles for each field, for protons and helium, and $10^7$ primary particles for each field for carbon.

\subsubsection{Monte Carlo configuration}
We use the "Dose To Medium" scorer in TOPAS to calculate the absorbed dose. To calculate cluster dose, we employed the same MCTS database and custom scorers as in \citeasnoun{faddegon_ionization_2023}. The scorers accumulate the track-weighted sum of the $I_p$: at each simulation step, the $I_p^c$ value is interpolated on-the-fly based on the particle class $c$, where the particle type and pre-step kinetic energy are used. This approach aligns with the default fluence filters implemented in TOPAS. Neutrons and particles with zero energy are excluded from the cluster dose scoring.

\subsubsection{Comparative analysis}
Dose and cluster dose distributions are compared using exemplary isocenter slices, depth profiles, and $\gamma$-analyses using a $\SI{2}{\percent}/\SI{2}{\milli\meter}$ distance-to-agreement criterion and a threshold of $10\%$ of the maximum dose. For the prostate patient, we also compute dose and cluster dose volume histograms. All metrics are computed with matRad.

\section{Results}
\subsection{Water box phantom: dose SOBP vs. cluster dose SOBP}\label{sec:3.1boxphantomPlans}

\begin{figure}[p]
    \centering
    \includegraphics[width=\textwidth]{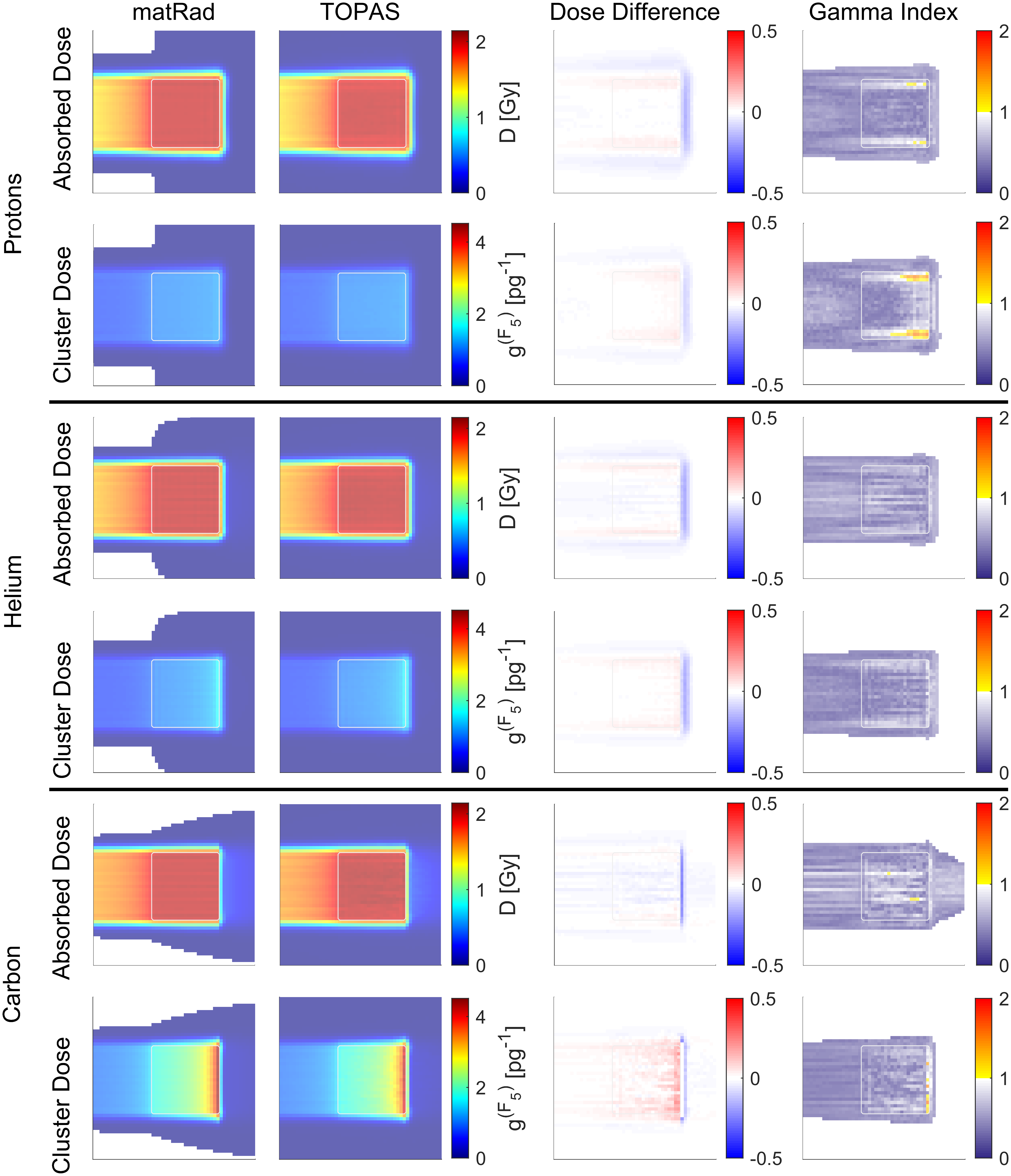}
    \caption{Absorbed dose and cluster dose $g^{(F_5)}$ distributions obtained with a prescribed absorbed dose SOBP, with the prescription $D=2 \si{\gray}$ within the target volume. Each row displays the absorbed dose (or cluster dose) distribution obtained with pencil-beam algorithm in matRad in the first panel, the corresponding result from MC in TOPAS in the second panel, the absolute difference of the two distributions is in the third panel and the Gamma Index in the fourth panel. Rows 1 and 2 are for protons, rows 3 and 4 for helium, and rows 5 and 6 for carbon.}
    \label{fig:BOXdoseOptiColor}
\end{figure}

Figure \ref{fig:BOXdoseOptiColor} shows the dose distributions for the central slice in the resulting treatment plans on the water box phantom with prescribed constant absorbed dose. For each of the three primary ion species we show the absorbed dose SOBP and the correspondent $g^{(F_5)}$ SOBP, We also show the results of the comparison of the pencil-beam kernel computation with the TOPAS computation by means of dose differences and $\gamma$-analyses.

\begin{figure}[p]
    \centering
    \includegraphics[width=\textwidth]{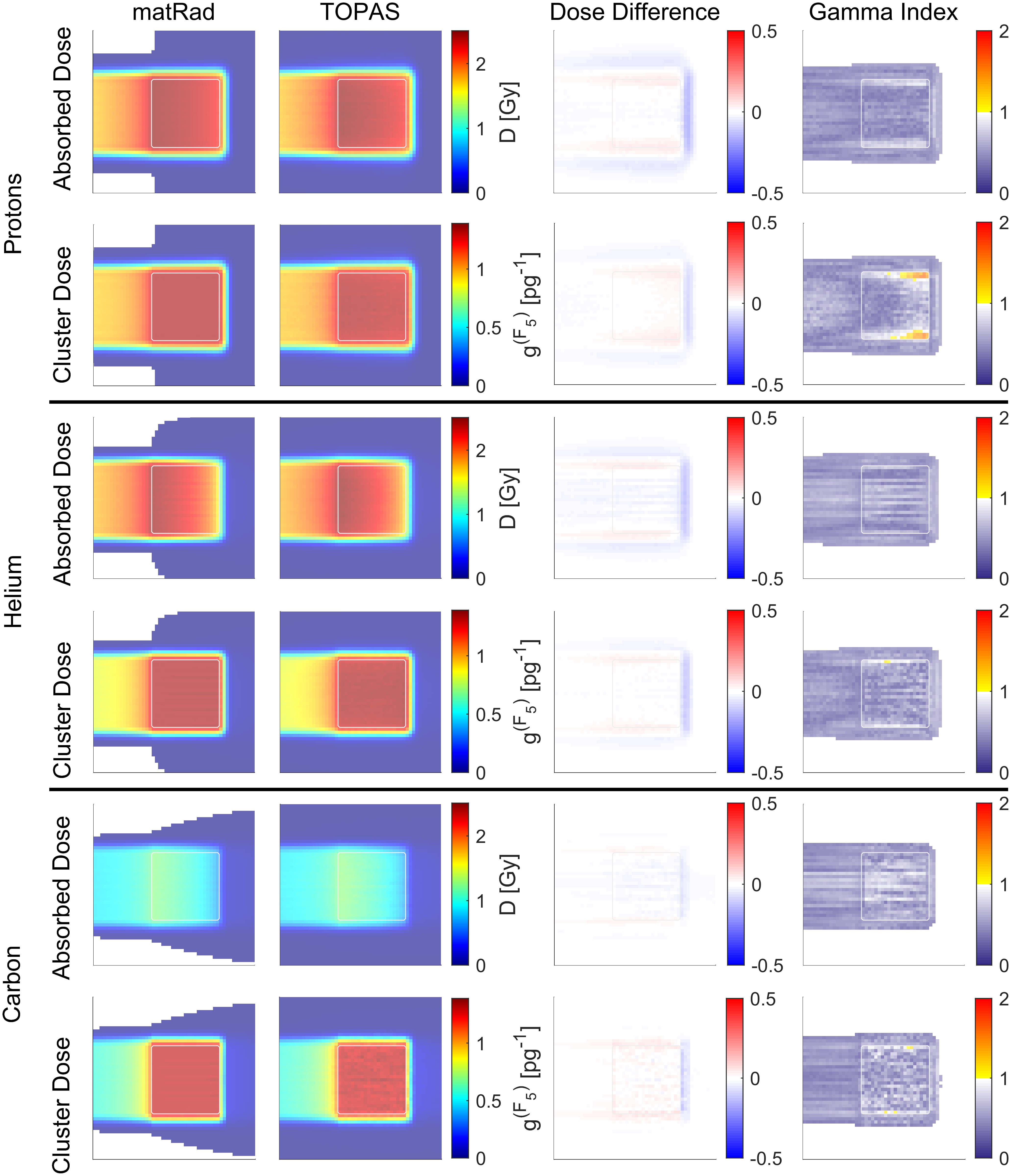} 
    \caption{Absorbed dose and cluster dose $g^{(F_5)}$ distributions obtained with a prescribed cluster dose $g^{(F_5)}$ SOBP, with the prescription $g^{(F_5)}= \SI{1.3}{\pico\gram}^{-1}$ within the target volume. Each row displays the absorbed dose (or cluster dose) distribution obtained with pencil-beam algorithm in matRad in the first panel, the corresponding result from MC in TOPAS in the second panel, the absolute difference of the two distributions is in the third panel and the Gamma Index in the fourth panel. Rows 1 and 2 are for protons, rows 3 and 4 for helium, and rows 5 and 6 for carbon.}
    \label{fig:BOXF5doseOptiColor}
\end{figure}

Figure (\ref{fig:BOXF5doseOptiColor}) shows the analogous results for a prescribed constant \textit{cluster dose} $g^{(F_5)}$, i.e.,  \enquote{cluster dose SOBP} in the boxphantom.

The absolute dose and cluster dose differences observed in Figures \ref{fig:BOXdoseOptiColor} and \ref{fig:BOXF5doseOptiColor} reveal a systematic lateral scattering inaccuracy most evident for protons, also substantiated by voxels outside of the $3\%/\SI{3}{\milli\meter}$ gamma criteria  in the computed $\gamma$ distributions. Since differences are consistent between cluster dose calculations and dose calculations, the differences are are attributed to a systematic error in lateral scattering. The $\gamma$ pass rates are all presented in Table \ref{GammaBox}.

\begin{table}
\caption{\label{GammaBox}Gamma analysis reflecting the comparison of RTP on a cubic water phantom with matRad vs. TOPAS recalculation. Two different prescriptions are applied for each irradiation mode: the first is a prescribed constant absorbed dose SOBP, and the second is a prescribed constant cluster dose SOBP. For each of the six RTP we compare two quantities, i.e. cluster dose $g^{(F_5)}$ and absorbed dose distributions. We list the GPR obtained with $2\%/2mm$ and $3\%/3mm$ distance-to-agreement criterium and threshold of $10\%$ of the maximum dose.}
\scriptsize
\resizebox{\textwidth}{!}{
\begin{tabular}{p{1.3cm}|  p{0.8cm} p{0.8cm} p{0.8cm} p{0.8cm} p{0.8cm} p{0.8cm} p{0.8cm} p{0.8cm} p{0.8cm} p{0.8cm} p{0.8cm} p{0.8cm}}
\br
&\centre{4}{Protons} &\centre{4}{Helium} &\centre{4}{Carbon}\\
\ns
&\crule{4}&\crule{4}&\crule{4}\\
& \multicolumn{2}{p{2.1cm}}{\centering Ab. Dose SOBP}& \multicolumn{2}{p{2.1cm}}{\centering Cluster Dose SOBP}& \multicolumn{2}{p{2.1cm}}{\centering Ab. Dose SOBP}& \multicolumn{2}{p{2.1cm}}{\centering Cluster Dose SOBP}& \multicolumn{2}{p{2.1cm}}{\centering Ab. Dose SOBP}& \multicolumn{2}{p{2.1cm}}{\centering Cluster Dose SOBP}\\
&\crule{4}&\crule{4}&\crule{4}\\
Gamma criteria& $D$ GPR & $g^{(F_5)}$ GPR & $D$ GPR & $g^{(F_5)}$ GPR & $D$ GPR & $g^{(F_5)}$ GPR & $D$ GPR & $g^{(F_5)}$ GPR & $D$ GPR & $g^{(F_5)}$ GPR & $D$ GPR & $g^{(F_5)}$ GPR\\
\mr
$2mm/2\%$& 96.2\% & 93.6\% & 97.3\% & 93.8\% & 98.5\% & 97.3\% & 99.3\% & 97.1\% & 97.6\% & 98.5\% & 98.1\% & 97.4\%\\
$3mm/3\%$& 99.0\% & 97.1\% & 99.3\% & 97.1\% & 99.9\% & 99.8\% & $>$99.9\% & 99.3\% & 99.9\% & 99.7\% & $>$99.9\% & 99.8\%\\
\br
\end{tabular}}
\end{table}
 
We can identify regions where the $\gamma$-analysis is outside of the criteria inside the target, near to the target boundaries and outside the target. But as the gamma passing rates demonstrate, the number of such voxels is small compared to the total. In all the plans on the boxphantom, the Gamma Pass Rate values for absorbed Dose are greater than $96.2\%$, while the values calculated on cluster dose distributions are greater than $93.6\% $. The GPRs of cluster dose appear to be smaller than the ones of absorbed dose.

\begin{figure}
    \centering
    \includegraphics[]{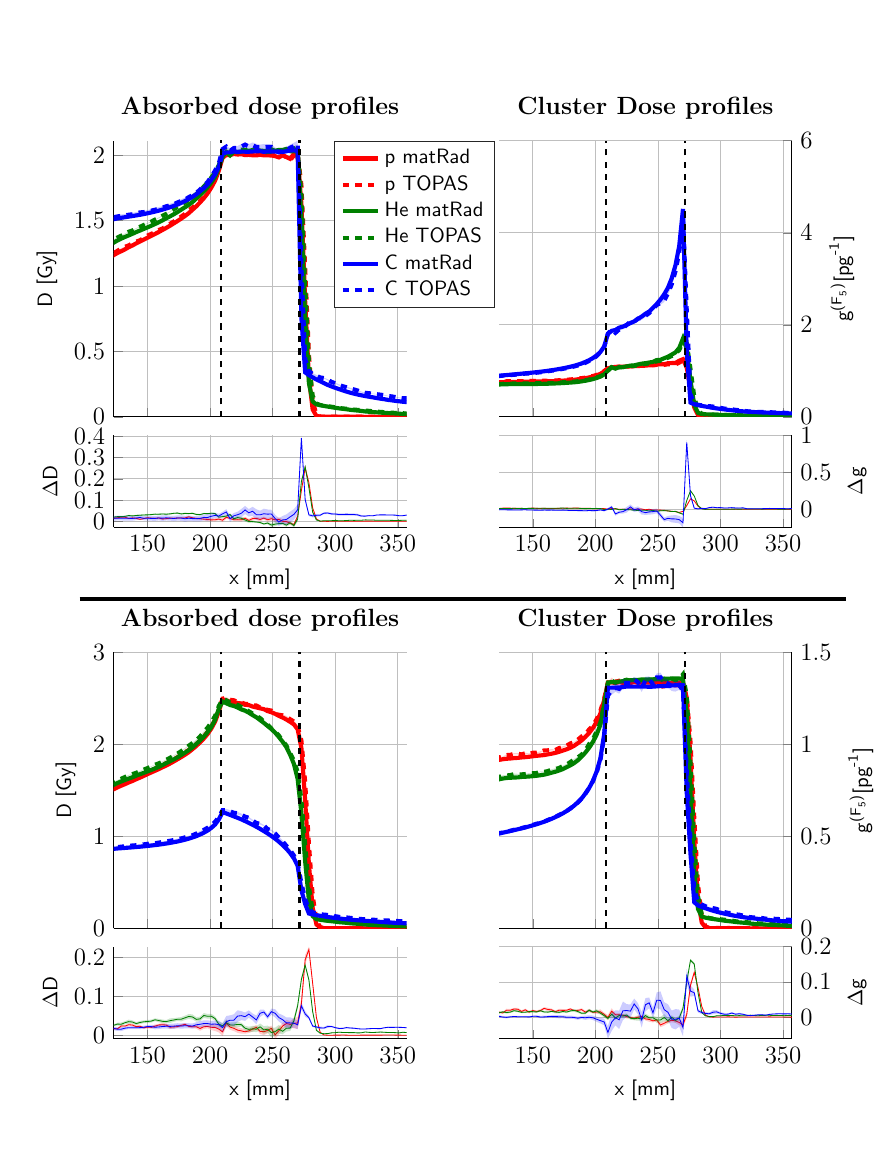}
    \caption{Dose profiles (left column) and cluster dose $g^{(F_5)}$ profiles (right column) for protons (red), helium (green) and carbon ions (blue). The results obtained with pencil-beam in matRad (solid line) are plotted against the results obtained with MC in TOPAS (dashed line). Results for absorbed dose SOBP optimization (first row) are separated from cluster dose SOBP optimization (second row). Below each profile plot the absolute dose difference profiles and error (shaded area) are shown.}
\label{fig:profiles_box_doseopti}
\end{figure}

Figure \ref{fig:profiles_box_doseopti} displays the profiles in depth along the central axis for absorbed dose and $g^{(F_5)}$ from the slice distributions in Figures \ref{fig:BOXdoseOptiColor} and \ref{fig:BOXF5doseOptiColor}. It also presents the profiles of the absolute dose differences between MC and PB, along with the corresponding standard deviation. The largest differences sistematically occur at the distal edge of the field, which corresponds to high-gradient regions in the dose distributions.

The results in Figures \ref{fig:BOXdoseOptiColor}, \ref{fig:BOXF5doseOptiColor} and \ref{fig:profiles_box_doseopti} confirm the expected behavior of cluster dose in depth. When simulating a constant dose SOBP, the cluster dose increases towards the distal edge. The magnitude and steepness of this increase depends on the used primary particle, as carbon ions create ionization clusters with a much higher frequency than protons. When optimizing a constant cluster dose SOBP, the physical dose needs to be reduced towards the distal edge.

In the next section we also quantify the expected improvement in cluster dose calculation obtained using the flexible method, i.e. taking into account the lateral fluence distributions for each primary and secondary particle type (see Section \ref{sec:flexiCalc}).

\subsection{Flexible cluster dose calculation}\label{sec:flexiCalc}

\begin{figure}[b]
    \centering
    \includegraphics[width=\textwidth]{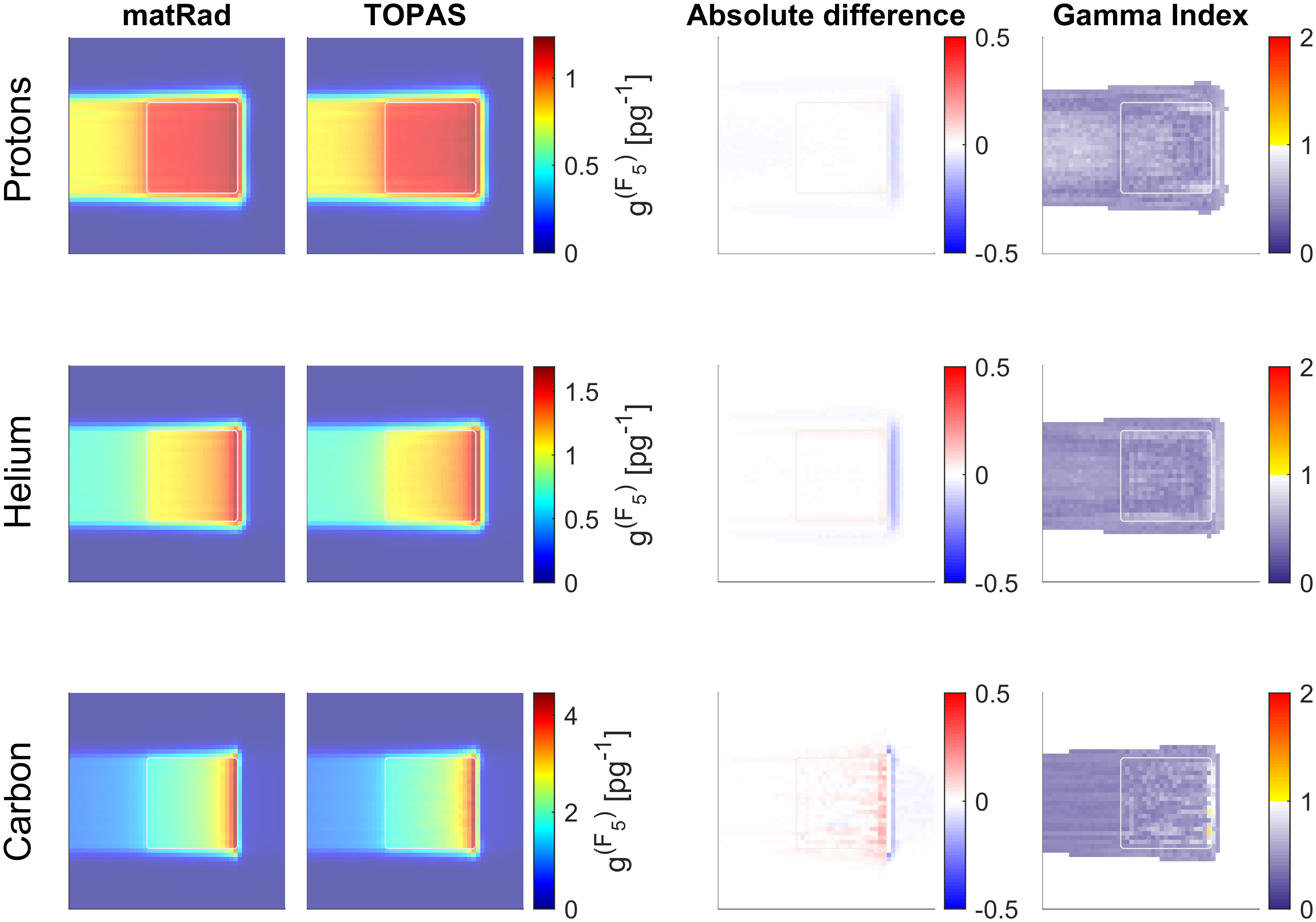} 
    \caption{Cluster dose $g^{(F_5)}$ calculated for an absorbed dose SOBP in the box phantom using the fluence kernel spectra in matRad (first column) is compared to the TOPAS recalculation (second column). Absolute difference (third column) and Gamma Index (fourth column) are included. First row is for protons, second row is for helium and third row is for carbon.}
    \label{fig:BoxFlexi}
\end{figure}

In this section we show the recalculation of cluster dose using the flexible method described in Section \ref{section2:clusterDoseCalc}, employing full fluence kernels spectra. We consider the previously optimized beam fluence from absorbed dose SOBPs in a boxphantom (see Figure \ref{fig:BOXdoseOptiColor}), and recalculate the corresponding cluster dose $g^{(F_5)}$. Results are displayed in Figure \ref{fig:BoxFlexi}, where we chose the same dose slices as in Figure \ref{fig:BOXdoseOptiColor} for comparison.

The absolute differences in the cluster dose slices obtained using the flexible method are smaller -- in particular, for protons the magnitude is half -- if compared to the fast cluster dose calculation method.

The improvement can be also assessed from the $\gamma$-analysis, where the Gamma Index failure regions present in the lateral target margins are reduced.

By comparing with the same MC $g^{(F5)}$ distributions displayed in Section \ref{sec:3.1boxphantomPlans}, the proton cluster dose benefits the most in terms of improvement, as GPR increases by $\sim 5.7\%$ to $GPR = 99.3\%$. Helium GPR increases by $\sim 2.3\%$ to $GPR=99.6\%$. Carbon GPR increases by $0.7\%$ to $GPR=99.2\%$.

\subsection{Calculation of cluster dose of different $I_p$}\label{sec:multipleIp}
\begin{figure}
    \centering
    \includegraphics[]{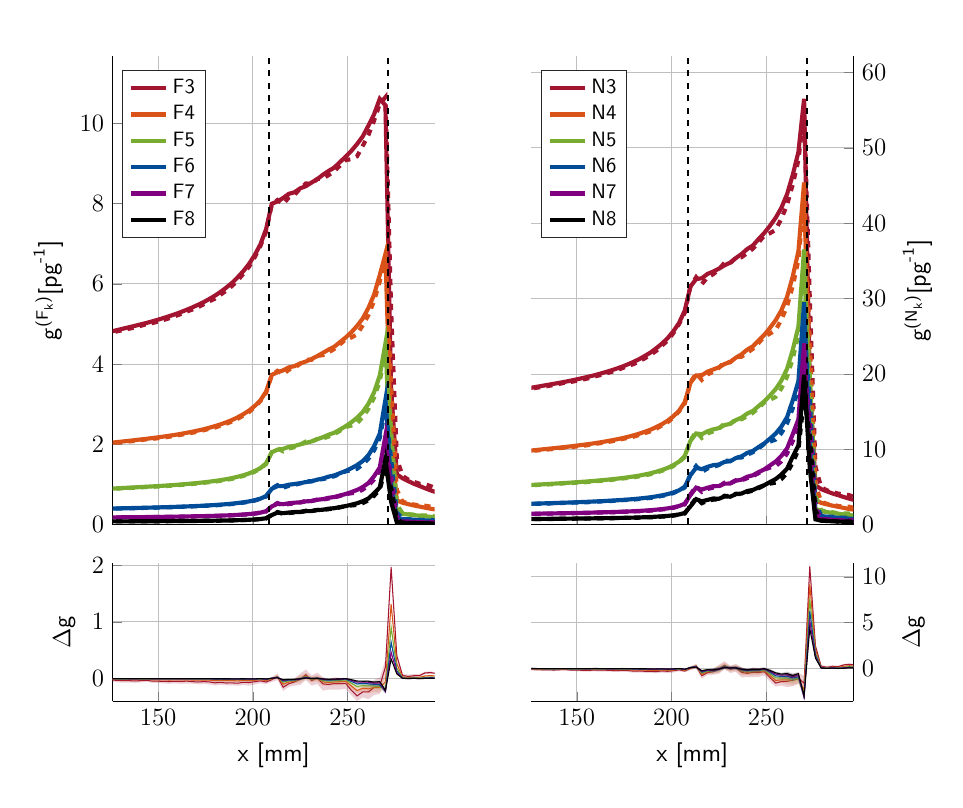}
    \caption{Cluster dose profiles determined by the carbon absorbed dose SOBP, for different $F_k$ (left panel) and $N_k$ (right panel), computed with matRad's pencil beam algorithms (solid line) and TOPAS (dashed line). Below each profile plot the absolute dose difference profiles and error (shaded area) are shown.}
    \label{fig:BoxDifferentIPs}
\end{figure}

Figure \ref{fig:BoxDifferentIPs} displays superimposed profiles of $g^{(F_k)}$, $g^{(N_k)}$ for each $k=\{3, ..., 8\}$, which were produced by the carbon absorbed dose SOBP within a boxphantom, calculated with the technique used in \ref{sec:3.1boxphantomPlans}, i.e. the \textit{fast} method for cluster dose calculation. 

The results show a consistent expected increase of cluster dose towards the distal edge. Cluster dose values as well as cluster frequencies decrease for larger $k$, and for all $k$ we observe $N_k > F_k$ which both follow by definition of the $I_p$.
The ratio between the peak value and the value at the target entrance is larger for larger $k$, consistent with the expectation of larger ionization clusters concentrating in the distal edge of the SOBP. All profiles exhibit good agreement between pencil-beam calculations and MC simulation.

\subsection{Prostate patient: Absorbed dose vs. Cluster dose optimization}
\begin{figure}
    \centering
    \includegraphics[width=\textwidth]{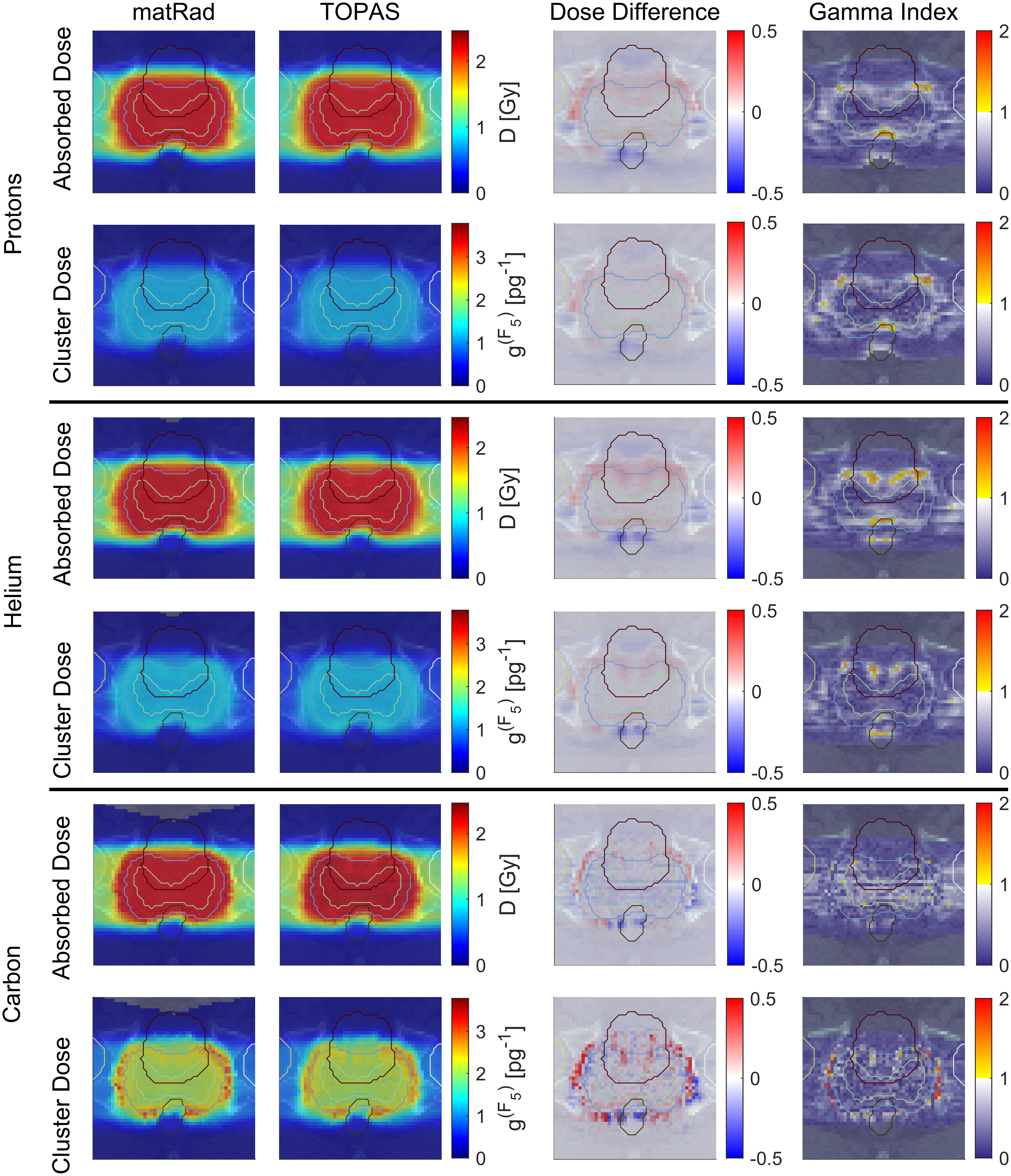} 
    \caption{Absorbed dose and cluster dose $g^{(F_5)}$ distributions obtained with a prescribed absorbed dose $\SI{2.27}{\gray}$ in the PTV prostate. Each row displays the absorbed dose (or cluster dose) distribution obtained with pencil-beam algorithm in matRad in the first panel, the corresponding result from MC in TOPAS in the second panel, the absolute difference of the two distributions is in the third panel and the Gamma Index in the fourth panel. Rows 1 and 2 show protons, rows 3 and 4 helium, and rows 5 and 6 carbon ions.}
    \label{fig:ProstatedoseOptiColor}
\end{figure}

The dose and $g^{(F_5)}$ slice distributions obtained with absorbed dose optimization on the PTV prostate are displayed in Figure (\ref{fig:ProstatedoseOptiColor}). Again, we compare the pencil-beam results to the MC results by means of absolute difference and $\gamma$-analysis.

\begin{figure}
    \centering
    \includegraphics{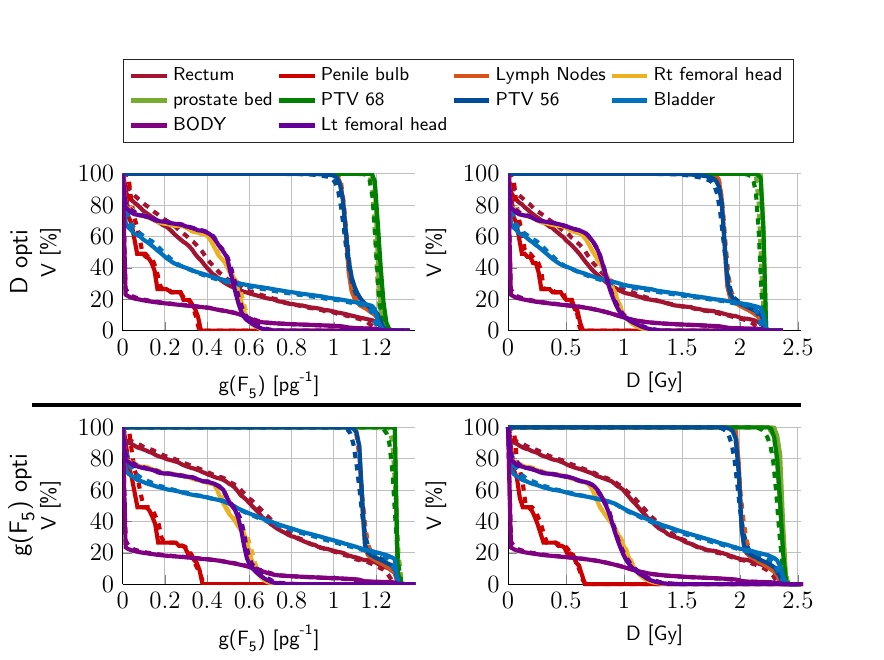}
    \caption{Dose volume histogram for protons prostate plans, comparing matRad PB (solid) vs. TOPAS (dashed). First row shows the results from absorbed dose optimization in the PTV. Second row shows the corresponding results for cluster dose $g^{(F_5)}$ optimization.}
    \label{fig:Prostate_dvh_protons}
\end{figure}

\begin{figure}
    \centering
    \includegraphics{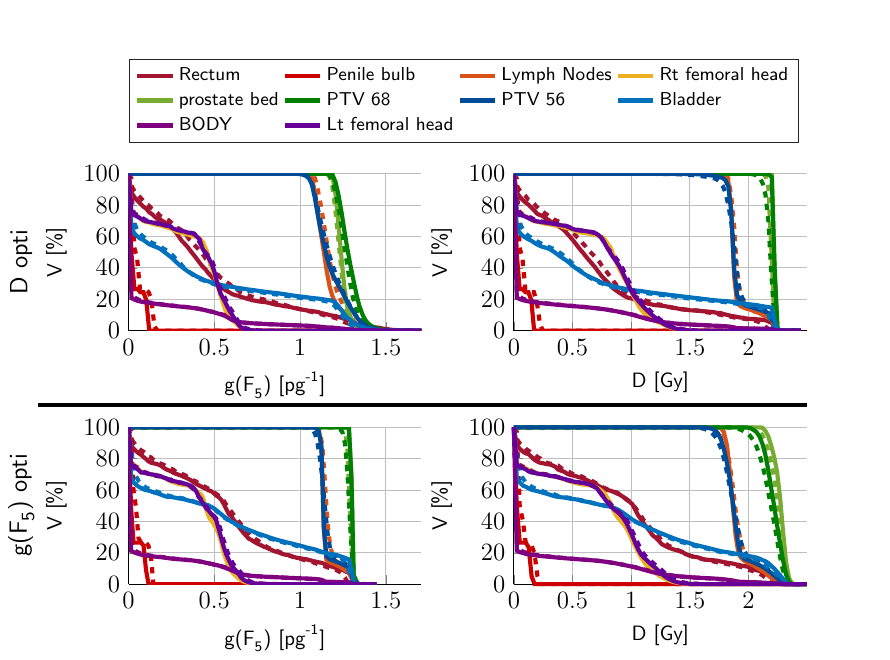}
    \caption{Dose volume histogram for helium prostate plans, comparing matRad PB (solid) vs. TOPAS (dashed). First row shows the results from absorbed dose optimization in the PTV. Second row shows the corresponding results for cluster dose $g^{(F_5)}$ optimization.}
    \label{fig:Prostate_dvh_helium}
\end{figure}

\begin{figure}
    \centering
    \includegraphics{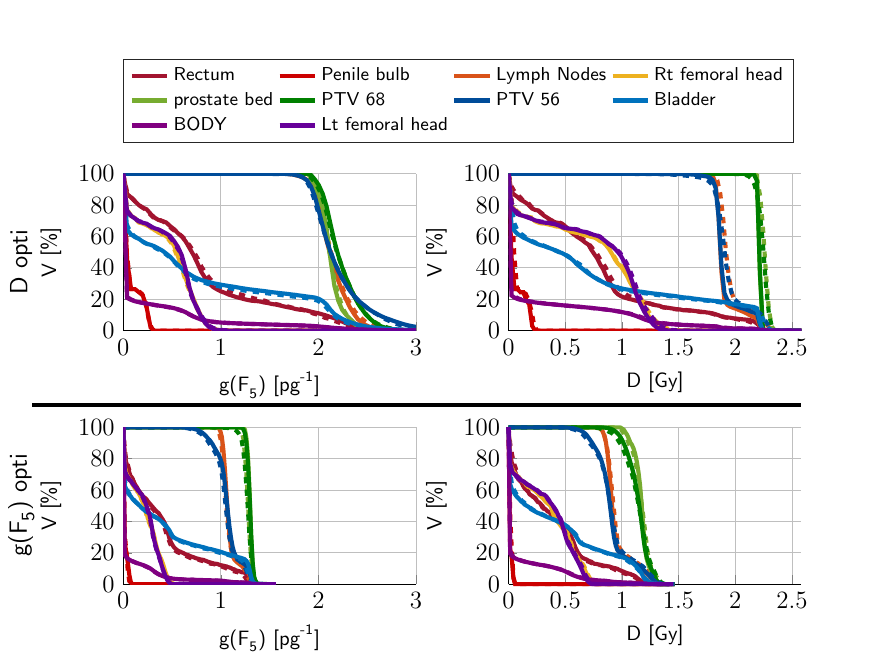}
    \caption{Dose volume histogram for carbon prostate plans, comparing matRad PB (solid) vs. TOPAS (dashed). First row shows the results from absorbed dose optimization in the PTV. Second row shows the corresponding results for cluster dose $g^{(F_5)}$ optimization.}
    \label{fig:Prostate_dvh_carbon}
\end{figure}

The absorbed dose optimization in the PTV prostate (Figure \ref{fig:ProstatedoseOptiColor}) demonstrates that protons show the most homogeneous $g^{(F_5)}$ within the target volume, if compared to helium and carbon (see also the Dose Volume Histograms in Figures \ref{fig:Prostate_dvh_protons}, \ref{fig:Prostate_dvh_helium}, and \ref{fig:Prostate_dvh_carbon}). This is consistent with the $g^{(F_5)}$ profiles previously shown for the absorbed dose SOBP in the box phantom (see Figure (\ref{fig:profiles_box_doseopti})).

We also applied our tool for prescribing a constant cluster dose level in the PTV prostate. While the prescription of constant cluster dose is based on the expected same biological effect for same $I_p$ and same cluster dose level (see \citeasnoun{faddegon_ionization_2023}), the appropriate values for cluster dose levels and constraints are still to be explored. But in this study, in order to choose a cluster dose level that is also reasonably consistent with the absorbed dose levels prescribed in clinics, we extrapolated the cluster dose level from a proton plan with prescribed constant dose in the PTV, assuming a constant biological effectiveness of RBE = 1.1. Thus cluster dose of $\bar{g}^{(F_5)} =\SI{1.3}{\per\pico\gram}$ in the PTV prostate was used as a reasonable prescription to optimize the corresponding cluster dose plans for protons, helium and carbon ions.
 
\begin{figure}
    \centering
    \includegraphics[width=\textwidth]{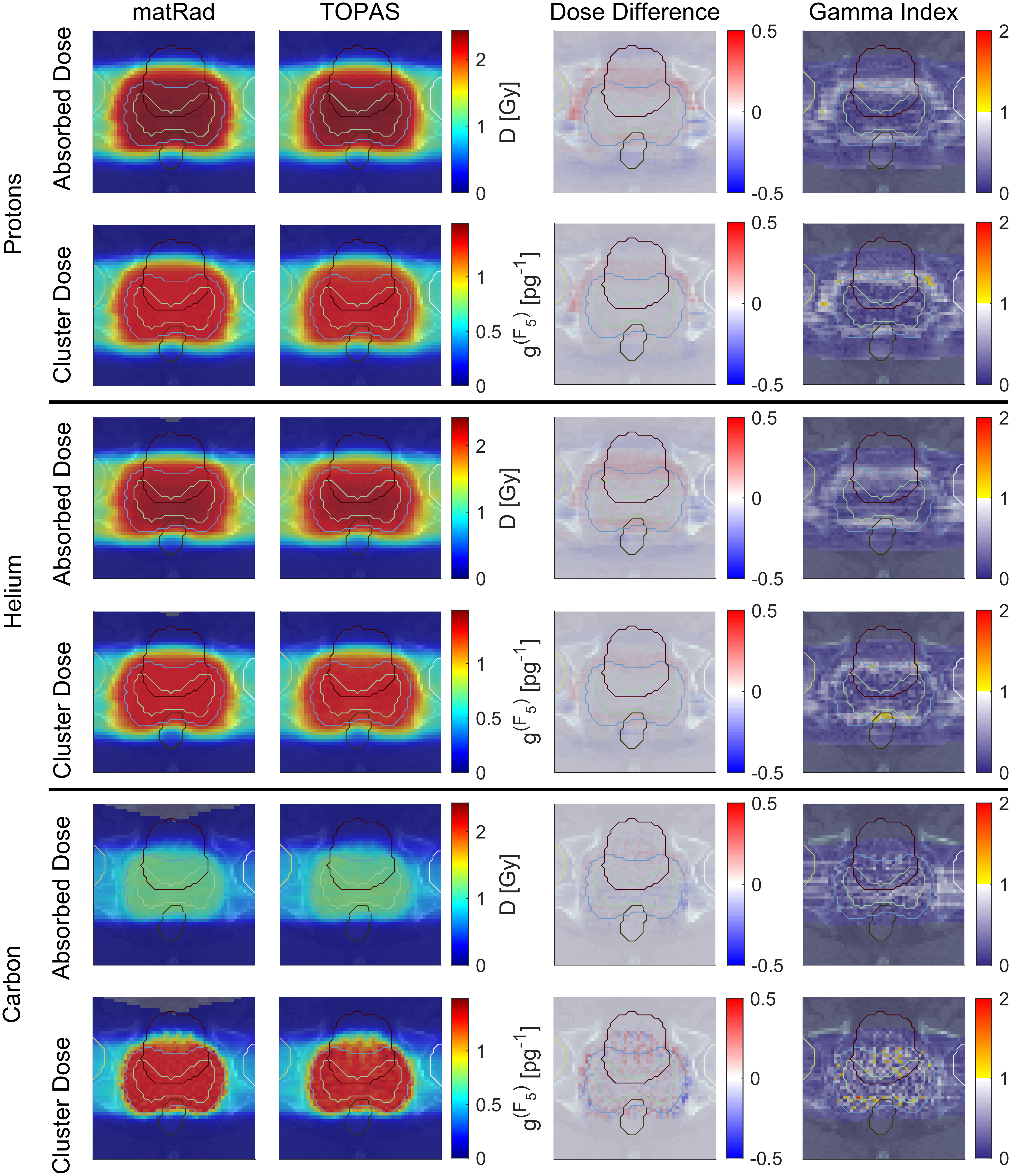} 
    \caption{Absorbed dose and cluster dose $g^{(F_5)}$ distributions obtained with a prescribed cluster dose $g^{(F_5)} = \SI{1.3}{\pico\gram}^{-1}$ in the PTV prostate. Each row displays the absorbed dose (or cluster dose) distribution obtained with pencil-beam algorithm in matRad in the first panel, the corresponding result from MC in TOPAS in the second panel, the absolute difference of the two distributions is in the third panel and the Gamma Index in the fourth panel. Rows 1 and 2 are for protons, rows 3 and 4 for helium, and rows 5 and 6 for carbon.}
    \label{fig:ProstateF5doseOptiColor}
\end{figure}

The corresponding dose and $g^{(F_5)}$ distributions are displayed in Figure (\ref{fig:ProstateF5doseOptiColor}).

From the prostate plans results we learn that a prescribed constant absorbed dose in the PTV determines increasing cluster dose in the distal edges, and thus peaks of cluster dose at the outer target boundary. This behavior is similar to the boxphantom results.

The (cluster) dose difference maps show the largest differences appearing near to the target's outer boundary, i.e., regions with high dose gradients. We can also observe that such boundary regions are characterized by a systematic dose overestimation in the regions intersecting the bladder, and dose underestimation in the regions intersecting the rectum. 

The relevant regions where the $\gamma$-analysis is outside the criteria are identified in the proximity of high dose gradients, as for the absolute differences. Other failure regions, visible in the PTV in carbon plans may also stem from lack of sufficient statistics of the MC simulation.
Considering the total pass-rates in Table \ref{GammaProstate}, we observe values of GPR greater than $91.3\%$ for dose and $94.4\%$ for cluster dose. 

\begin{table}[bh]
\caption{\label{GammaProstate}Gamma analysis reflecting the comparison of RTP on a prostate patient with matRad vs. TOPAS recalculation. For each irradiation mode we compare the two different optimizations, or plans, and for each of them we compare two quantities, i.e. cluster dose $g^{(F_5)}$ and absorbed dose distributions.}
\scriptsize
\resizebox{\textwidth}{!}{
\begin{tabular}{p{1.3cm}|  p{0.8cm} p{0.8cm} p{0.8cm} p{0.8cm} p{0.8cm} p{0.8cm} p{0.8cm} p{0.8cm} p{0.8cm} p{0.8cm} p{0.8cm} p{0.8cm}}
\br
&\centre{4}{Protons} &\centre{4}{Helium} &\centre{4}{Carbon}\\
\ns
&\crule{4}&\crule{4}&\crule{4}\\
& \multicolumn{2}{p{2.1cm}}{\centering Ab. Dose Optimization}& \multicolumn{2}{p{2.1cm}}{\centering Cluster Dose Optimization}& \multicolumn{2}{p{2.1cm}}{\centering Ab. Dose Optimization}& \multicolumn{2}{p{2.1cm}}{\centering Cluster Dose Optimization}& \multicolumn{2}{p{2.1cm}}{\centering Ab. Dose Optimization}& \multicolumn{2}{p{2.1cm}}{\centering Cluster Dose Optimization}\\
&\crule{4}&\crule{4}&\crule{4}\\
Gamma criteria& $D$ GPR & $g^{(F_5)}$ GPR & $D$ GPR & $g^{(F_5)}$ GPR & $D$ GPR & $g^{(F_5)}$ GPR & $D$ GPR & $g^{(F_5)}$ GPR & $D$ GPR & $g^{(F_5)}$ GPR & $D$ GPR & $g^{(F_5)}$ GPR\\
\mr
$2mm/2\%$& 95.6\% & 94.8\% & 96.2\% & 94.8\% & 93.5\% & 95.4\% & 94.5\% & 94.4\% & 91.3\% & 96.0\% & 96.9\% & 95.2\%\\
$3mm/3\%$& 99.1\% & 98.8\% & 99.2\% & 98.8\% & 98.4\% & 99.0\% & 99.0\% & 98.9\% & 98.8\% & 99.0\% & 99.6\% & 98.7\%\\
\br
\end{tabular}
}
\end{table}

Figure \ref{fig:Prostate_dvh_protons} shows the Dose Volume Histograms (DVHs) from proton plans. First, it is possible to see that optimizing on absorbed dose determines a rather homogeneous cluster dose and vice versa. 

In PTV, the matRad PB (cluster) dose overestimates the TOPAS MC results: the $D_{95}$ for PTV56 and PTV68 differs by maximum 3.2\%. Considering the analogous quality factor for cluster dose, that we can call $G_{95}$ (minimum cluster dose received by 95\% of target volume), we obtain a maximum deviation of 3.4\%, that is comparable to the absorbed dose deviation. Regarding the maximum deviations in the mean doses in the two PTVs, for mean absorbed dose is 1.6\% and for mean cluster dose is 1.5\%. 
Regarding the OARs, in the bladder the mean dose differs by maximum 0.4\%, while the mean cluster dose differs by maximum 0.07\%. In the Rectum, the mean dose is underestimated by maximum -5.2\%, and similarly the mean cluster dose by maximum -5.4\%.

Figure \ref{fig:Prostate_dvh_helium} shows the Dose Volume Histograms from helium plans. In PTVs, $D_{95}$ differs by maximum 4.5\%, and $G_{95}$ differs by maximum 3.5\%. The mean dose differs by maximum 1.9\%, and the mean cluster dose differs by maximum 1.5\%. In OARs, the maximum differences in mean doses are visible in Rectum, with -5.2\% for the mean absorbed dose and -6.6\% for mean cluster dose. In the bladder, the mean dose differs by maximum 0.01\% and the mean cluster dose differs by maximum 0.07\%.

Figure \ref{fig:Prostate_dvh_carbon} shows the Dose Volume Histograms from carbon plans. In PTVs, $D_{95}$ differs by maximum 4.6\%, and $G_{95}$ analogously differs by maximum 4.7\%. Comparing mean values in the PTVs, the maximum deivations are 1.4\% for mean absorbed dose, and  2\% for mean cluster dose. In OARs, the maximum deviations in the mean dose is -2\% observed in the Rectum. Similarly, the maximum deviation in the mean cluster dose is 3.4\% in the Rectum. In the bladder, the mean dose differs by maximum 1.4\%, while the mean cluster dose differs by maximum 3\%.

\subsection{Comparison of calculation times}
The rationale of this work is to enable fast and direct cluster dose calculation using PB algorithm, since full transport MC is currently unfeasible due to its high computational cost. We presented and validated two methods for cluster dose calculation using PB algorithm, namely the \textit{fast} and the \textit{flexible} methods, as already described in Section \ref{sec:gcalcConcept}.

In this section, we provide the computation times for both methods. Table \ref{tab:dijcalctime} reports the calculation times for the influence matrix across the three radiation modalities explored and the different phantoms. This refers to the stage before fluence optimization, as the latter does not depend on the method used to calculate the $d_{ij}$ and $g_{ij}$. As shown in the table, in general, the flexible method leads to an increase in the calculation times of about one order of magnitude. This increase is more pronounced for carbon, as a larger set of particle types is considered in the kernel database.

\begin{table}[H]
\begin{center}

\caption{\label{tab:dijcalctime} Calculation times (in seconds) of dose influence matrix for fast and flexible methods. Different radiation modalities and phantom geometries are included.}
\scriptsize
\resizebox{0.5\textwidth}{!}{
\begin{tabular}{p{1.3cm} p{2cm}  p{1.5cm}  p{1.5cm} }
\br
&& Fast &Flexible\\
\mr
Protons & Boxphantom & 28s & 254s \\
& \crule{3}\\
& Prostate & 151s & 1203s \\
\mr
Helium & Boxphantom & 27s & 254s\\
& \crule{3}\\
& Prostate & 201s & 1438s \\
\mr
Carbon & Boxphantom & 24s & 576s \\
& \crule{3}\\
& Prostate & 114s & 1719s \\
\br
\end{tabular}
}
    
\end{center}
\end{table}

\section{Discussion}
\subsection{Summary}
We demonstrated a workflow for inclusion and direct optimization of nanodosimetric ID in RTP with protons, helium and carbon ions. The tool employs an extended version of pencil-beam algorithm in matRad, that relies on precomputed particle fluence kernels in water to calculate the pencil-beam cluster dose. We provided two methods for the cluster dose pencil-beam calculation, one faster with a higher degree of approximation, and the other one slower and flexible, which directly uses the fluence kernel data. 

\subsection{Pencil-beam cluster dose calculation accuracy}
The first question we aimed to answer was the degree of reliability in using PB algorithm for RTP with ID. This was assessed in a validation of the implemented PB-algorithms against TOPAS, calculating 12 different RTP, testing different patient geometries, radiation modalities and prescriptions on cluster dose and absorbed dose. For demonstration purposes we chose $F_5$, i.e. the preferred $I_p$ for normoxic cells \cite{faddegon_ionization_2023}.

In a boxphantom, the limited lateral scattering model of cluster dose kernels in the fast method results in the largest differences to MC recalculation. This becomes evident when using the flexible method, i.e., the full fluence-spectra kernels which, as expected, provides results in better agreement with MC. The improvement over the fast method is more evident for protons, considering a more strict criterium for the gamma analysis of $2mm/2\%$ criterium and threshold $10\%$ of the maximum dose, showing up to $5\%$ improvement of the GPR values. If we consider $3mm/3\%$ criterium and threshold $10\%$ of the maximum dose, the improvement is of up to $3\%$. The improvement over the fast method is limited if we consider helium and carbon ions, considering the larger kernel database and computation time. 

In the prostate phantom, the main dose and cluster dose differences are visible in regions of lateral cluster dose gradients, substantiating above suspicion of lateral scattering limitations as major contributor to cluster dose differences. Nevertheless, using the fast method we obtained acceptable $GPRs > 98.4\% $ using the $3mm/3\%$ criterium and threshold $10\%$ of the maximum dose. Analyzing the DVHs, however, this lateral systematic error manifests in reduced coverage in the MC recalculated plans. This reduction is consistent between dose and cluster dose, allowing the conclusion that reducing the inherent systematic error (e.g., through pencil-beam subsampling) would improve both cluster dose and dose calculations and that no substantial methodological component hinders pencil-beam cluster dose calculations compared to conventional dose calculation.

\subsection{Cluster dose planning}
Since cluster-dose essentially provides a fluence-dependent macroscopic quantity measuring the deposition of ionization clusters in matter, we hypothesized its usability for direct treatment planning. Since cluster dose prescriptions are unknown, we explored an approach to derive a cluster-dose prescription from a conventional proton treatment plan optimized on dose, where the radiobiological effectiveness may be assumed close to constant within the target volume.  

Applying the derived prescription to ions heavier than protons results in behavior expected from the cluster dose recalculations on dose optimized plans, and also general experience from RBE-weighted dose planning \cite{karger_rbe_2021,frese_mechanism-based_2012} and LET distributions \cite{durante_physics_2021,glowa_relative_2024} for ions. Use of heavier ions leads to lower physical dose required, which in our work is due to their much higher frequency of producing nanoscopic ionization clusters. 
This behavior is both visible for the cubic phantom irradiated with one field and for the prostate phantom irradiated with two opposing beams.

\subsection{Scope and limitations}
While our work provides a framework for cluster dose treatment planning based on ID within acceptable accuracy margins, assumptions and numerical, technical details may limit its scope. For example, the fluence spectra kernels can take substantial size, depending on the choice of the energy and depth binning, and thus trade accuracy against computational complexity. We deem a thorough investigation of this aspect out of scope, as we suspect behavior consistent with using similar spectra for RBE-modeling, for example. Further, typical for a pencil-beam kernel algorithm, the kernels are pre-computed in water, and differences of ID in the various media encountered within the human body are so far ignored. However, this limitation translates to the availability of MCTS data and is also mitigated by the fact that water still makes up the predominant medium in the cell nucleus.

Another aspect to put under scrutiny is the role of secondary electrons (i.e., $\delta$-rays) and keeping their \enquote{count} consistent between MCTS simulations and macroscopic planning tools. This work includes electrons implicitly considered in the ID database, meaning that electrons within a range of approximately \SI{100}{\nano\meter} from the primary track are accounted for. This, however, excludes $\delta$-rays with a larger range, which could lead to an underestimation of the cluster dose as their ID is not explicitly considered.

Yet, we argue that future changes to the inclusion of $\delta$-rays would not compromise the validity of our framework as long as handling of $\delta$-rays is consistent between the MC simulation and the pencil-beam kernels. 
The electron $I_p$ could then be included when scoring $\delta$-rays with large range in the kernel precomputation as well as MC simulations.

\subsection{Future Outlook}
As this work fundamentally introduced MC validated cluster dose planning, the natural next step would be a comprehensive side-by-side comparison of RBE-weighted planning to cluster-dose planning. Concurrently, dual planning approaches might be of interest, where cluster dose or voxel-average ID steers a clinically acceptable RBE-weighted planning approach towards a favorable distribution of ionization clusters, already proposed by \cite{burigo_simultaneous_2019,yang_nanodosimetric_2024}. Here, it would be natural to connect to existing work on LET-driven planning \cite{grassberger_variations_2011,giantsoudi_linear_2013,unkelbach_reoptimization_2016} and multi-ion therapy seeking homogeneous dose-weighted LET distributions \cite{ebner_emerging_2021}. 
A more distant, futuristic goal may finally be primarily cluster-dose prescribed treatment plans, capturing biological response through personalized and local response models, as has already been proposed based on LET.

\section{Conclusion}
This work presents a MC validated cluster dose treatment planning framework within matRad for protons, helium and carbon ions. Using prototypical plans on a box phantom and a prostate patient showed that established pencil-beam algorithms for cluster dose calculation yield typical pencil-beam accuracy compared to MC simulations in TOPAS. 
 
Consequently, our framework provides a fast and valid cluster dose planning for nanodosimetric research, while retaining flexibility for custom track structure databases and/or other ID parameters.

The framework may enable future work on a comparative study with RBE-weighted dose planning, direct planning on voxel-averaged $I_p$, and validating direct MC planning with fast MC codes.

\section{Acknowledgements}
This research was supported by NIH grant R01 CA266467. I would like to acknowledge Dr. José Ramos Méndez for the valuable discussions with our collaborators, which provided insights that informed fundamental aspects of our work.

\section*{References}
\bibliographystyle{jphysicsB}
\bibliography{BibliographyPaper1}

\end{document}